\documentclass{ws-mpla}
\usepackage{graphicx}
\usepackage[super,compress]{cite}
\usepackage{tensor}
\usepackage{hyperref}
\usepackage{url}
\hypersetup{colorlinks,urlcolor=cyan,citecolor=green,linkcolor=blue,filecolor=black}
\begin{document}

\catchline{}{}{}{}{}
\title{L\'{e}vy-Leblond Equation and Eisenhart-Duval lift in Koopman-von Neumann Mechanics}

\author{Bikram Keshari Parida}
\address{Sun Moon University, South Korea\\ parida.bikram90.bkp@gmail.com}

\author{Abhijit Sen}
\address{Novosibirsk State University, Russia and \\ 
Tulane University, USA.\\
abhijit913@gmail.com}

\author{Shailesh Dhasmana}
\address{Novosibirsk State University, Russia\\ dx.shailesh@gmail.com}
 
\author{Zurab K. Silagadze}
\address{Budker Institute of Nuclear Physics and \\ Novosibirsk State University, 630 090, Novosibirsk, Russia. \\ silagadze@inp.nsk.su}

\maketitle

\pub{Received (Day Month Year)}{Revised (Day Month Year)}
 
\begin{abstract}
The Koopman-von Neumann (KvN) mechanics is an approach that was formulated long ago to answer the question regarding the existence of a Hilbert space representation of classical mechanics. KvN mechanics is a non-relativistic theory, and it is not clear how spin can be included in it, since spin is widely regarded as a relativistic property. In Eur. Phys. J. Spec. Top. {\bf 227}, 2195 (2019) \cite{cabrera_2019} it was argued that the Spohn equation [Ann. Phys. {\bf 282}, 420 (2000)] \cite{spohn_2000} is the correct classical framework for the Koopman-von Neumann theory corresponding to the Dirac equation. However, after L\'{e}vy-Leblond's seminal work on this topic, it became clear that spin naturally arises also from the Galilean invariant wave equations, without any need of relativistic considerations. Inspired by this, we propose another possibility of including spin in the KvN formalism: the L\'{e}vy-Leblond equation in the Koopman-von Neumann formalism can be obtained as a null reduction of the massless Dirac equation in the Eisenhart-Duval lift metric. To illustrate the idea, we implement it for a one-dimensional classical system without magnetic interactions.
\keywords{L\'{e}vy-Leblond Equation; Eisenhart-Duval lift; Koopman-von Neumann Mechanics.}
\end{abstract}

\section{Introduction}
The Hilbert space formulation of classical mechanics was the subject of research a long time ago, shortly after the quantum mechanics entered the physical literature with its representation in Hilbert space. Although the quantum world differs from the classical world both conceptually and in the mathematical formalism used, there should be a smooth transition from one world to another. The search for a common framework for both quantum mechanics and classical mechanics led to the creation of the Koopman-von Neumann mechanics \cite{kvn_1931,neuman_1932}. However, at that time KvN formulation failed to attract due attention, and only recently we are witnessing a renewed interest in it from the research community, which has led to significant contributions in this direction  (see for example \cite{Gozzi:1986ge,Gozzi:1989bf,Mauro_2003,bondar_2019,Gay_2020,bermudez_2020,Bermudez_2021,Cattaruzza:2018epp,Gay-Balmaz:2021zve,Morgan:2019azd,McCaul:2022cyl,Tronci:2022csw} for some of them). 

The KvN formalism of non-relativistic classical mechanics is the most natural in the phase space. However, introducing auxiliary operators, this theory can be re-formulated in the configuration space too, where the KvN mechanics becomes more quantum-like. Moreover, Sudarshan's perspective \cite{sudarshan_1976,Sudarshan_1979,chashchina_2020,sen_2022} allows to consider the classical subspace as a part of twice larger genuinely quantum system with specific dynamics. Interestingly, quantum-mechanics-free subsystem, introduced by Tsang and Caves in Ref. \refcite{Tsang_2012} and even experimentally realized \cite{Mercier_de_L_pinay_2021}, is nothing but the realization of Sudarshan's perspective on the Koopman-von Neumann mechanics \cite{Silagadze_2023}.

Quite recently,  KvN mechanics was geometrized \cite{sen_2022} by using the Eisenhart lift toolkit \cite{eisenhart_1928,Cariglia_2015,dhasmana_2021}. The Eisenhart lift, rediscovered by Duval \cite{Duval_1985,Duval_1991}, establishes an interesting connection between non-relativistic dynamics on the Galilean manifold and light-like geodesics in Minkowski spacetime. In the case of KvN dynamics, the Minkowski spacetime is replaced by a pseudo-Riemannian manifold of signature $(2,2)$ (Kleinian spacetime with ultra-hyperbolic geometry) \cite{sen_2022}. We extend this geometric perspective on the KvN mechanics by considering the massless Dirac equation in the Eisenhart-Duval lift metric. Having performed its null reduction, we will find an equation of L\'{e}vy-Leblond type in the case of KvN mechanics.

The L\'evy-Leblond equation \cite{levy_1967,Wilkes_2020} was a significant landmark in the history of spin \cite{Tomonaga_1998}. The Klien-Gordon (KG) equation (an attempt to create a relativistic quantum mechanics) was riddled with problems, such as negative probabilities when interpreted as a one-particle wave equation. To get rid of these problems, Dirac proposed an equation that is linear in both space and time derivatives. The Dirac equation naturally introduces the concept of spin, and therefore spin was considered as a relativistic phenomenon. However, it was realized by L\'{e}vy-Leblond that Schr\"odinger equation (spin-0 non-relativistic equation) can be viewed as a non-relativistic equivalent of KG equation and a linearized version of the Schr\"odinger equation can produce spin without any mention of relativity \cite{levy_1967, Kasri2011}.

In the section that follows, we will employ the Eisenhart-Duval lift technique to derive the L\'{e}vy-Leblond equation in the KvN case for a one-dimensional classical system without magnetic interactions, rather than linearizing the corresponding Koopman-von Neumann equation.

Throughout the article natural units with $\hbar=1$, $c=1$ are assumed.

\section{L\'evy- Leblond equation in KvN case from a lightlike reduction}

Null-reduction of the massless KG equation on the base space of the Eisenhart-Duval spacetime
yields the Schr\"{o}dinger equation \cite{Horvathy_2009}. The same formalism makes it possible to unify relativistic and non-relativistic quantum theories of spin systems \cite{Duval_1987}. Namely, L\'{e}vy-Leblond equation can be obtained by null reduction of the massless Dirac equation on the base space of the Eisenhart-Duval spacetime \cite{Duval_1996,Cariglia_2012}.

The same program can be undertaken in the case of KvN mechanics. In Ref. \refcite{sen_2022} Eisenhart-Duval lift of the Koopman-von Neumann mechanics was considered. The resulting ultra-hyperbolic metric has the form
\begin{align}
    dS^2_{KvN} = 2 dQ \, dq + 2 ds \, dt - \dfrac{2}{m} Q \dfrac{\partial V(q,t)}{\partial t} \, dt^2.
    \label{eq1}
\end{align}
As in the case of quantum mechanics, the KvN equation can be considered as a null reduction of the massless KG equation in this metric \cite{sen_2022}. Now it's time to see what happens in the case of the massless Dirac equation, which in this metric has the form \cite{parker_2009}
\begin{align}
    i \; \tensor{e}{_a ^\mu} \;\gamma^{a} \; \nabla_{\mu} \Psi = 0, 
    \label{direq}
\end{align}
where the co-variant derivatives of the Dirac spinor field $\Psi (Q,q,s,t)$ is 
\begin{align}
    \nabla_{\mu} \Psi = \partial _{\mu} \Psi + i \; \tensor{\Omega}{_\mu _a _b } \; \Sigma^{ab} \Psi, \label{covarder}
\end{align}
with $\Psi$ being the 4-component wave function
\begin{align}
    \Psi (Q,q,s,t) = \left( \begin{array}{c}
        \psi_{1}(Q,q,s,t) \\
          \psi_{2}(Q,q,s,t)\\
          \psi_{3}(Q,q,s,t)\\
          \psi_{4}(Q,q,s,t)
    \end{array} \right). \label{psisp}
\end{align}
 Note that we will use Latin indices for the tensor components in the local orthonormal frame (locally inertial frame), in which the metric tensor is
\begin{align}
    \eta_{ab} = 
     \left( 
    \begin{array}{cccc}
1 & \quad 0 & \quad 0 & \quad 0\\
0 & \quad 1 & \quad 0 & \quad 0\\
0 & \quad 0 & \quad -1 & \quad 0\\
0 & 0 & 0 & -1
    \end{array}
    \right), 
    \label{eq3}
\end{align}
while, according to (\ref{eq1}), in the original coordinates $(Q,q,s,t)$ the components of the metric are
\begin{align}
    g_{\mu \nu} = \left( 
    \begin{array}{cccc}
0 & \quad 1 & \quad 0 & \quad 0\\
1 & \quad 0 & \quad 0 & \quad 0\\
0 & \quad 0 & \quad 0 & \quad 1\\
0 & \quad 0 & \quad 1 & \quad - \dfrac{2 Q}{m} \dfrac{\partial V(q,t)}{\partial q}
    \end{array}
    \right),\;\;\;
   g^{\mu \nu} = \left( 
    \begin{array}{cccc}
0 & \quad 1 & \quad 0 & \quad 0\\
1 & \quad 0 & \quad 0 & \quad 0\\
0 & \quad 0 & \quad \dfrac{2 Q}{m} \dfrac{\partial V(q,t)}{\partial q} & \quad 1\\
0 & \quad 0 & \quad 1 & \quad 0
    \end{array}
    \right).   
    \label{eq2}
\end{align}
In Eq.(\ref{covarder}),
\begin{align}
    \Sigma^{ab} = - \dfrac{i}{8} \; \left[\gamma^{a} , \gamma^{b} \right],
\end{align}
and the Christoffel symbols and spin-connection coefficients are given by the usual expressions 
\begin{align}
  &  \Gamma^{\alpha}_{\; \mu \nu } = \dfrac{1}{2} g^{\alpha \beta} \left( \partial_{\mu} g_{\beta \nu} + \partial_{\nu} g_{\beta \mu} - \partial_{\beta} g_{\mu \nu }\right),
 &   \tensor{\Omega}{_\mu ^a _b} = e_{b}^{\; \rho} \; e^{a}_{\; \nu} \; \Gamma^{\nu}_{\; \mu \rho} - e_{b}^{\; \nu} \; \partial_{\mu} e^{a}_{\; \nu}.
 \label{Chr-Spin}
\end{align}

The vierbein $e_{a} ^{\,\, \mu}$ is the matrix that transforms between the tetrad frame and the coordinate frame. It satisfies $e_{a} ^{\,\, \mu} g_{\mu \nu} e_{b} ^{\,\, \nu} = \eta_{ab}$ and is defined up to a $SO(2,2)$ transformation. In our case, the vierbein can be chosen in the form
\begin{align}
    e_{a} ^{\, \, \mu} = \left(
    \begin{array}{cccc}
      0    & \quad 0 & \quad \dfrac{1}{2} + \dfrac{Q}{m} \dfrac{\partial V(q,t )}{\partial q }\quad & 1  \\
       -1   & \quad -\dfrac{1}{2} & 0 & 0 \\
       0 & \quad 0 & \dfrac{1}{2} - \dfrac{Q}{m} \dfrac{\partial V(q,t) }{\partial q } & -1 \\
       1 & \quad -\dfrac{1 }{2} & 0&0
    \end{array}\right) ,
    \label{tet1}
\end{align}
its inverse being
\begin{align}
    e^{a}_{\,\, \mu} = \left(
    \begin{array}{cccc}
     0    & \quad 0 & \quad 1 &\quad \dfrac{1}{2} - \dfrac{Q}{m} \dfrac{\partial V(q,t)}{\partial q } \\
     - \dfrac{1}{2}    & \quad -1 & \quad 0 & 0\\
     0 & \quad 0 & \quad 1 & \quad - \dfrac{1}{2} - \dfrac{Q}{m} \dfrac{\partial V(q,t) }{\partial q } \\
     \dfrac{1}{2} &\quad  -1 & \quad 0 & 0
    \end{array}
    \right). 
    \label{tet2}
\end{align}
We verified the correctness of the  above expressions by checking that 
\begin{align*}
    \nabla _{\mu} e^{a}_{\,\, \nu} = \partial _{\mu} e^{a}_{\,\, \nu} + \tensor{\Omega}{_\mu ^a _b} \,\,  e^{b}_{\; \nu} - \Gamma^{\rho}_{\; \mu \nu } \; e^{a}_{\; \rho} = 0.
\end{align*}

Finally, constant gamma matrices in the tetrad frame satisfying the Clifford algebra 
\begin{equation}
\gamma^{a} \gamma^{b} + \gamma^{b} \gamma^{a} = 2 \, \eta^{ab}
\label{Clifford}
\end{equation}
can be given as 
\begin{align}
 &   \gamma^{1} = \left(
    \begin{array}{cccc}
    0 & \quad 0 & \quad 0 & \quad 1\\
    0 & \quad 0 & \quad1 & \quad 0 \\
    0 & \quad 1 & \quad 0 & \quad 0 \\
    1 & \quad 0 & \quad 0 & \quad 0
    \end{array}\right) ,  \quad \quad  & \gamma^{2} = \left(
    \begin{array}{cccc}
   0 & \quad 0 & \quad 1 & \quad 0\\
   0 & \quad 0 & \quad 0 & \quad -1 \\
   1 & \quad 0 & \quad 0 & \quad 0\\
   0 & \quad -1 & \quad 0  & \quad 0
    \end{array}\right) , \nonumber \\
 &   \gamma^{3} =  \left(
    \begin{array}{cccc}
   0 & \quad 0 & \quad 0 & \quad -1 \\
   0 & \quad  0 & \quad 1 & \quad 0\\
   0 & \quad -1 & \quad 0 & \quad 0\\
   1 & \quad 0 & \quad 0 & \quad 0 
    \end{array}\right), \quad \quad & \gamma^{4} =  \left(
    \begin{array}{cccc}
   0 & \quad 0 & \quad -1 & \quad 0 \\
   0 & \quad 0 & \quad 0 & \quad -1 \\
   1 & \quad 0 & \quad 0 & \quad 0 \\
   0 & \quad 1 & \quad 0 & \quad 0
    \end{array}\right). \label{gammamet}
\end{align}
Non-zero Christoffel symbols and spin-connection coefficients are calculated as 
\begin{align}
& \Gamma^q_{\,tt}=-\Gamma^s_{\,Qt}=-\Gamma^s_{\,tQ}=\frac{1}{m}\frac{\partial V}{\partial q},
\nonumber \\
& \Gamma^Q_{\,tt}=-\Gamma^s_{\,qt}=-\Gamma^s_{\,tq}=\frac{Q}{m}\frac{\partial^2 V}{\partial q^2},\;\;\Gamma^s_{\,tt}=-\frac{Q}{m}\frac{\partial^2 V}{\partial q\partial t},
\end{align}
and
\begin{align}
 &\Omega_{t12}=-\Omega_{t21}=\Omega_{t23}=-\Omega_{t32}=\frac{1}{m}\frac{\partial V}{\partial q}+\frac{Q}{2m}\frac{\partial^2 V}{\partial q^2}, \nonumber \\
 &\Omega_{t34}=-\Omega_{t43}=\Omega_{t41}=-\Omega_{t14}=\frac{1}{m}\frac{\partial V}{\partial q}-\frac{Q}{2m}\frac{\partial^2 V}{\partial q^2}.
\end{align}
Accordingly, by virtue of \eqref{Clifford} we obtain
\begin{align}
i \, \Omega_{tab}\Sigma^{ab}=\frac{1}{2}(\gamma^1-\gamma^3)(\Omega_{t12}\gamma^2-\Omega_{t34}\gamma^4).
\end{align}
On the other hand, $e_a^{\;t}\gamma^a=\gamma^1-\gamma^3$, and since according to \eqref{Clifford}
$(\gamma^1-\gamma^3)^2=0$, we see that the spin connection does not contribute to the Dirac equation \eqref{direq}. As a result, from the Dirac equation \eqref{direq} we obtain the following system of equations
\begin{align}
 &   -2 \dfrac{\partial \psi_{3}}{\partial Q} + 2 \left( \dfrac{\partial \psi_{4}}{\partial t} + \dfrac{Q}{m} \dfrac{\partial V(q,t)}{\partial q} \dfrac{\partial \psi_{4}}{\partial s}  \right)=0, \;\;\;
    \dfrac{\partial \psi_{3}}{\partial s} + \dfrac{\partial \psi_{4}}{\partial q}=0, \nonumber\\
 &   - \dfrac{\partial \psi_{1}}{\partial q} + 2 \left(\dfrac{\partial \psi_{2}}{\partial t} + \dfrac{Q}{m} \dfrac{\partial V(q,t)}{\partial q} \dfrac{\partial \psi_{2}}{\partial s} \right) =0, \;\;\;
  \;  \dfrac{\partial \psi_{1}}{\partial s} + 2 \dfrac{\partial \psi_{2}}{\partial Q}= 0.  \label{direq4}
\end{align}
The null reduction of the Dirac equation \eqref{direq} is achieved by requiring
\begin{align}
    \Psi (Q,q,s,t) = \Phi(Q,q,t) e^{ims}
    =\left(\begin{array}{c}
        \phi_{1}(Q,q,t)\\
         \phi_{2}(Q,q,t)\\
         \phi_{3}(Q,q,t)\\
         \phi_{4}(Q,q,t)
    \end{array} \right) \; e^{ims} .\label{psisp0}
\end{align}
Then we obtain from (\ref{direq4})
\begin{align}
&    - \dfrac{\partial \phi_{3}}{\partial Q} + \dfrac{\partial \phi_{4}}{\partial t} + i \, Q \dfrac{\partial V (q,t)}{\partial q } \phi_{4}= 0, \;\;\; \quad\quad
    \phi_{3} = \dfrac{i}{m} \dfrac{\partial \phi_{4}}{\partial q}, \nonumber \\                  
&    - \dfrac{\partial \phi_{1}}{\partial q} + 2 \left(\dfrac{\partial \phi_{2}}{\partial t} + i \, Q \dfrac{\partial V (q,t)}{\partial q}  \phi_{2}\right)= 0, \;\;\;
   \phi_{1} = \dfrac{2 i}{m} \dfrac{\partial \phi_{2}}{\partial Q}. \label{eq29a}
\end{align}
If we eliminate the non-dynamical components $\phi_1$ and $\phi_3$ using the second and fourth equations of this system, we find that the dynamical components $\phi_2$ and $\phi_4$ of the wave function both satisfy the KvN equation for a particle with mass $m$ and potential $V(q,t)$:
\begin{align}
&    i \dfrac{\partial \phi_{4}}{\partial t} = \left(Q \dfrac{\partial V(q,t)}{\partial q} -  \dfrac{1}{m} \dfrac{\partial^{2}}{\partial Q \partial q} \right) \phi_{4},\nonumber \\
&    i \dfrac{\partial \phi_{2}}{\partial t} = \left(Q \dfrac{\partial V(q,t)}{\partial q} -  \dfrac{1}{m} \dfrac{\partial^{2}}{\partial q \partial Q} \right) \phi_{2}. \label{kvn2}
\end{align}
Introducing two-component spinors $\chi = \left(\begin{array}{c} \phi_{1} \\ \phi_{3} \end{array} \right)$ and $ \xi = \left(\begin{array}{c} \phi_{4} \\ \phi_{2} \end{array} \right) $, as well as differential operators $\hat{p} = -i \dfrac{\partial  }{\partial Q}$ , $\hat{P} = -i \dfrac{\partial}{\partial q }$ and $\hat{E} = i \dfrac{\partial}{\partial t}$, system (\ref{eq29a}) takes the form of matrix equations:
\begin{align}
\left( \begin{array}{cc} 0   & \hat{p} \\ \frac{\hat{P}}{2}   & 0 
\end{array} \right) \chi + \hat{E} \; \xi - Q \dfrac{\partial V(q,t)}{\partial q} \xi =0,\;\;\;
\chi &= -\dfrac{2}{m} \left( \begin{array}{cc} 0   & \hat{p} \\ \dfrac{\hat{P}}{2}   & 0 \end{array}\right)  \xi. 
\label{eqn35a}
\end{align}
Finally, writing the matrix $\left( \begin{array}{cc}
      0   & \hat{p} \\
      \frac{\hat{P}}{2}   & 0 
    \end{array} \right)$ as a linear combination of the Pauli matrices 
\begin{align}
     \left( \begin{array}{cc}
      0   & \hat{p} \\
      \frac{\hat{P}}{2}   & 0 
    \end{array} \right) = \dfrac{\sigma_{1}}{2} \left[ \hat{p} + \dfrac{\hat{P}}{2} \right] + i\; \dfrac{\sigma_{2}}{2}  \left[ \hat{p} - \dfrac{\hat{P}}{2} \right],
    \end{align}
we obtain the L\'{e}vy-Leblond equation in the KvN case:
\begin{align}
    - \left[  \dfrac{\sigma_{1}}{2} \left( \hat{p} + \dfrac{\hat{P}}{2} \right) + i\; \dfrac{\sigma_{2}}{2}  \left( \hat{p} - \dfrac{\hat{P}}{2} \right) \right] \chi + Q \dfrac{\partial V(q,t)}{\partial q} \xi = \hat{E} \xi, \nonumber \\
    \chi = - \dfrac{1}{m} \left[ \sigma_{1} \left(\hat{p} + \dfrac{\hat{P}}{2} \right) + i \, \sigma_{2} \left( \hat{p} - \dfrac{\hat{P}}{2}\right)\right] \xi. \label{eq43a}
\end{align}
Another version of the KvN L\'evy-Leblond equation can be written in terms of $\sigma_{+}= \dfrac{\sigma_{1}+ i \, \sigma_{2}}{2}$ and $\sigma_{-} = \dfrac{\sigma_{1} - i \, \sigma_{2}}{2}$ matrices:
\begin{align}
    - \left[ \sigma_{+} \; \hat{p} + \sigma_{-} \; \dfrac{\hat{P}}{2}\right] \chi + Q \dfrac{\partial V(q,t)}{\partial q} \xi = \hat{E} \; \xi, \label{eq45a}\\
    \chi = -\dfrac{2}{m} \left[ \sigma_{+} \; \hat{p} + \sigma_{-} \; \dfrac{\hat{P}}{2}\right] \xi.\label{eq46a}
\end{align}
Like the Dirac equation, two spinors $\xi$ and $\chi$ can be unified in one bi-spinor $\left (\begin{array}{c} \xi \\ \chi \end{array}\right )$. However, unlike the Dirac case, the second spinor  $\chi$ is non-dynamical. The same is true for the L\'{e}vy-Leblond equation, and this is to be expected, since both the KvN equation and the Schr\"{o}dinger equation are of the first order in time, and L\'{e}vy-Leblond-type equations are their linearizations, a kind of ``square roots" of these equations \cite{Lazzarini_2020}.

\section{Conclusions}
A general holonomic conservative system in classical dynamics with $d$ degrees of freedom is geometrically described by the Eisenhart-Duval lift in terms of the geodesics of the Lorentzian metric in the $(d+2)$-dimensional space-time \cite{eisenhart_1928}. For treating time dependent dynamical systems and their symmetries, this geometric perspective is particularly convenient \cite{Cariglia_2018}.

The same geometric perspective provided by the Eisenhart-Duval lift can also be used in quantum theory, since null reduction of the massless KG equation from Eisenhart-Duval space-time leads to the Schr\"{o}dinger equation \cite{Horvathy_2009}, and a similar null reduction of the massless Dirac equation gives the L\'{e}vy-Leblond equation \cite{Duval_1996,Cariglia_2012}.

The Eisenhart-Duval toolkit can be applied to KvN mechanics as well and gives new and interesting insights into this theory \cite{sen_2022}. In this article, we extended this geometrical view on the KvN mechanics to the case of a non-relativistic spin described by the L\'{e}vy-Leblond equation.

We considered only the one-dimensional case. Physics in one dimension can seem rather boring. However, this is not at all true. In one dimension, an electron that is trying to propagate must push its neighbors. As a result, when interactions are switched on, individual motions are impossible and one has to consider collective excitations. On the other hand, the one-dimensional nature makes the problem simple enough that it can be solved almost completely analytically. As a result, one-dimensional systems of interacting particles, such as the Tomonaga-Luttinger liquid, have incredibly rich physics, with strange effects such as spin-charge separation that have fascinated both theorists and experimenters for more than 70 years \cite{Giamarchi_2003,Senaratne_2022}.

We do not know if the KvN L\'{e}vy-Leblond equation can find any practical application. However, it is remarkable that (two-dimensional) L\'{e}vy-Leblond equation is thoroughly used in geometric approach to low energy electronic properties of bilayer and few-layers graphene \cite{Cariglia_2017,Cariglia_2018a}. Another interesting application of the same  L\'{e}vy-Leblond equation can be found in Ref. \refcite{cariglia_2018b}. In this article L\'{e}vy-Leblond fermions are used in a study of entanglement generated in the $1+2$ geometry of a Bronnikov-Ellis wormhole. The most interesting thing is that a condensed matter analogue of this effect could be realized in the laboratory using two planes of two-layer graphene connected by a two-layer carbon nanotube \cite{cariglia_2018b}.

Nevertheless, a three-dimensional generalization of the results presented in this article is certainly of interest. For a KvN particle in three dimensions, the Eisenhart-Duval lift will result in an eight-dimensional ultra-hyperbolic space-time. Eight-dimensional space has distinctive characteristics that are unique to this dimensionality. Namely, the definition of what is a vector and what is spinors of the first and second kind in this space is only a matter of convention; all three quantities are perfectly equivalent \cite{Gamba_2004}. It will be interesting to find out whether this triality property of eight-dimensional space leaves any trace after null reduction in the corresponding KvN L\'{e}vy-Leblond equation.

Another deficiency of our approach is that it does not take into account magnetic interactions.
The introduction of electromagnetism in the case of KvN theory was studied in Ref. \refcite{Mauro_2003} and an interesting consequence is that there is no Aharonov-Bohm effect for the charged KvN particle, in contrast to the quantum case, since KvN theory is inherently classical in nature. The Eisenhart-Duval lift of a KvN mechanics with a non-zero vector potential requires special consideration, which is beyond the scope of this article.

\bibliographystyle{ws-mpla}
\bibliography{refkvn_LL}

\end{document}